\documentclass[final,3p,times]{elsarticle}

\usepackage{amssymb}
\usepackage{amsmath}
\usepackage{graphicx}
\usepackage{subcaption}
\usepackage{lineno}

\newcommand{\snn}{\sqrt{s_{_{\rm NN}}}}
\newcommand{\pt}{p_{\rm T}}
\newcommand{\mpt}{\langle p_{\rm T}\rangle}
\newcommand{\nch}{N_{\rm ch}}
\newcommand{\npart}{N_{\rm part}}

\newcommand{\cs}{c_s^2}
\newcommand{\jpsi}{J/\psi}
\newcommand{\OO}{O+O}
\newcommand{\dAu}{d+Au}
\newcommand{\AuAu}{Au+Au}
\newcommand{\lam}{\Lambda}
\newcommand{\lambar}{\bar{\Lambda}}

\journal{Journal of Subatomic Particles and Cosmology}

\begin{document}

\begin{frontmatter}

\title{STAR Highlights II: Study of Small Systems and the Search for New, Exotic Physics}

\author[sbu,bnl]{Jiangyong Jia (for the STAR Collaboration)}
\ead{jiangyong.jia@stonybrook.edu}
\affiliation[sbu]{organization={Department of Chemistry, Stony Brook University},
             city={Stony Brook},
             postcode={11794},
             state={NY},
             country={USA}}
\begin{abstract}
Twenty-five years of RHIC operation have produced a uniquely diverse dataset, which enables a broad physics program. This contribution highlights recent STAR results in four areas: exotic-state searches and ultra-peripheral collisions; the onset of quark--gluon plasma (QGP) signatures in small systems; radial flow and its fluctuations; and polarization and spin correlations. This proceedings summarizes the results presented in the Strangness in Quark Matter 2026 conference by the STAR collaboration.
\end{abstract}


\end{frontmatter}

\section{Introduction}
\label{sec:intro}
\begin{figure}[h]
\centering
\includegraphics[width=0.75\linewidth]{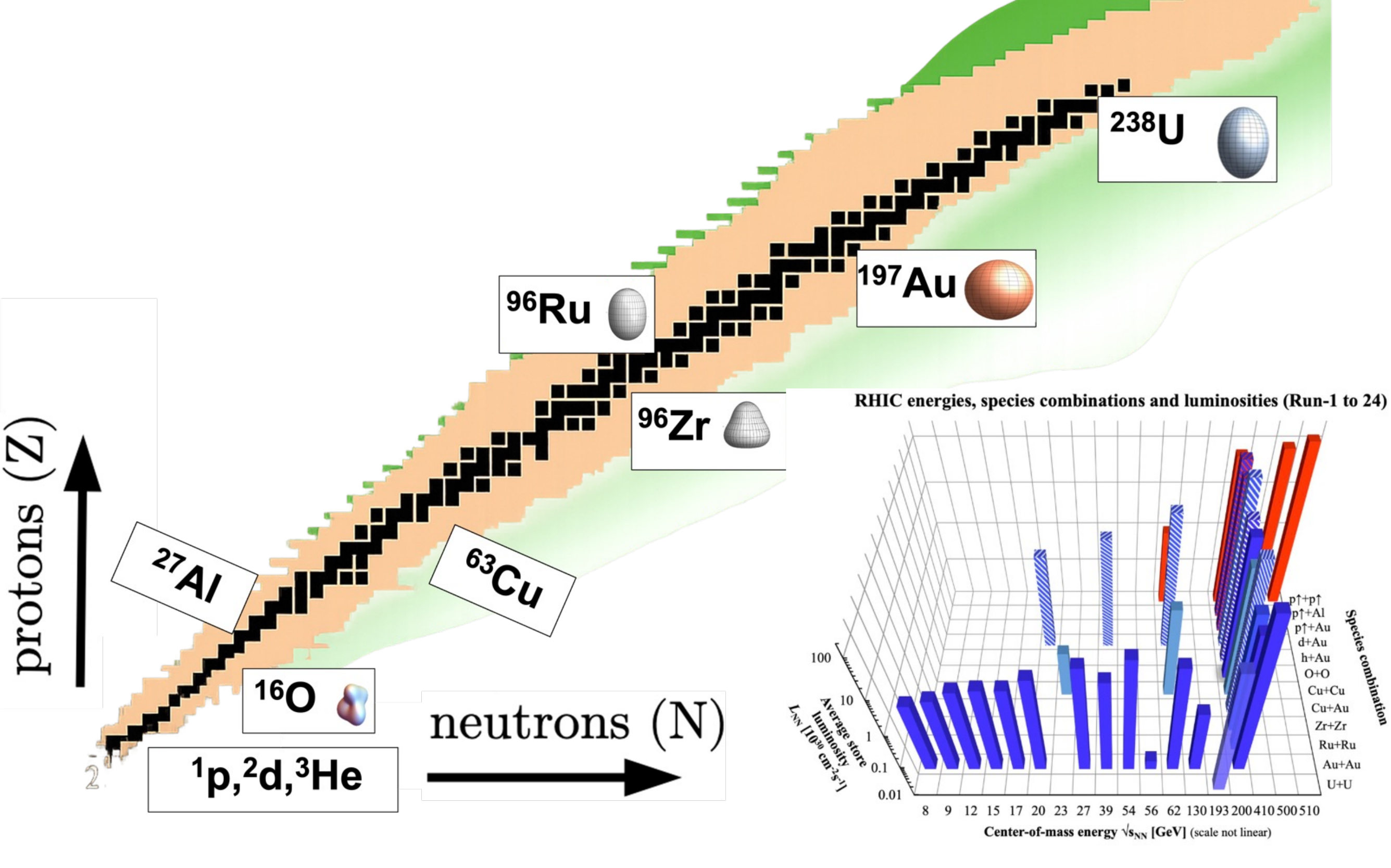}
\caption{The coverage of the STAR dataset in species and energy combinations.}
\label{fig:species}
\end{figure}

Twenty-five years of RHIC running have provided the STAR experiment with a dataset of exceptional breadth (Fig.~\ref{fig:species}): approximately thirty collision energies from $\snn=3$ GeV to 200 GeV in \AuAu\, and twelve unique species combinations from $pp$ and $p$+Au through $d$+Au, $^{3}$He+Au, \OO, Cu+Cu, the $^{96}$Ru+$^{96}$Ru/$^{96}$Zr+$^{96}$Zr isobar pair, \AuAu, and U+U. This diversity in system size, geometry, and energy is the foundation of a rich physics program that will continue to produce results for the next decade.

This contribution presents STAR highlights in four areas, following the structure of the conference talk: (i) exotic searches and ultra-peripheral collision (UPC) physics; (ii) the onset of quark--gluon plasma (QGP) signatures in \OO\ and \dAu\ collisions; (iii) radial flow and its fluctuations; and (iv) polarization and spin correlations. Each area is illustrated with representative measurements. Further details are given in the parallel talks and posters of the STAR contributors referenced throughout, who present their own proceedings in this volume. Unless otherwise noted, results labeled ``STAR Preliminary'' have not yet undergone final publication review.

\section{Exotic searches and UPC physics}
\label{sec:exotic}

\begin{figure}[h]
\centering
\includegraphics[width=1.01\linewidth]{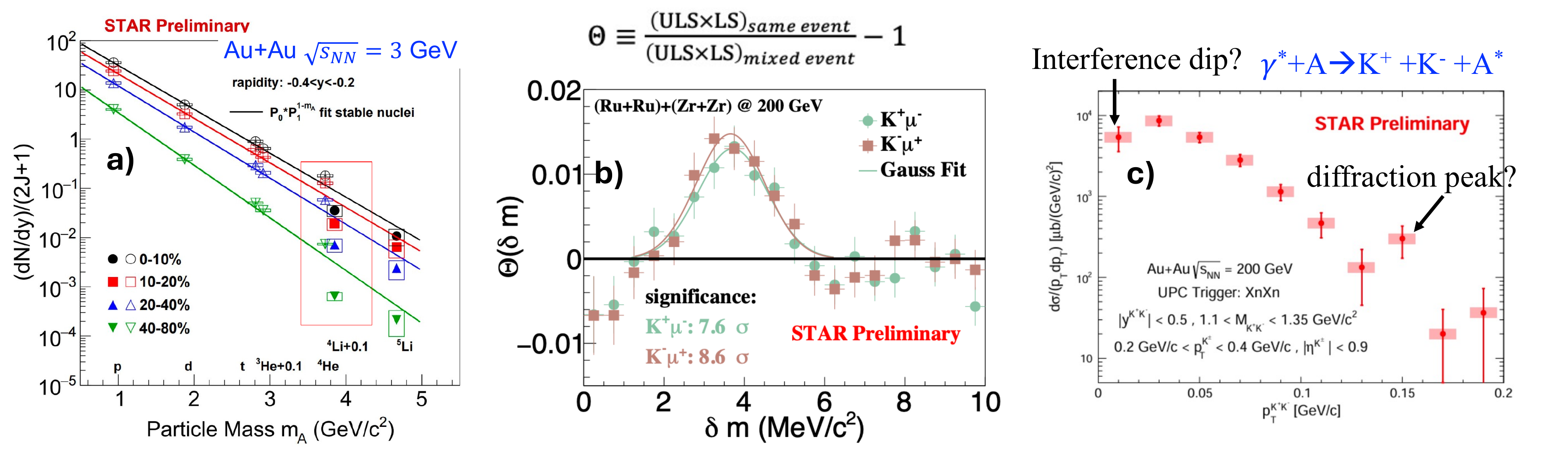}
\caption{(a) Yields of light nuclei per spin state, $(dN/dy)/(2J+1)$, as a function of mass number in \AuAu\ collisions at $\snn=3$~GeV for four centrality intervals~\cite{Wu:SQM2026}. Stable nuclei (open symbols) follow the exponential coalescence systematics (lines); the unstable $^{4}$Li and $^{5}$Li (filled symbols, boxed) fall below it. (b) Correlation strength $\Theta(\delta m)$ for $K^{+}\mu^{-}$ and $K^{-}\mu^{+}$ pairs as a function of the residual mass $\delta m = M_{\rm inv}-m_{\mu}-m_{K}$ in isobar collisions at $\snn=200$~GeV, showing the excess attributed to muonic-atom formation~\cite{XWang:SQM2026}. (c) Coherent non-resonant $K^{+}K^{-}$ photoproduction cross section as a function of pair $\pt$ in \AuAu\ UPCs at $\snn=200$~GeV, exhibiting an interference dip at low $\pt$ and a possible diffraction peak~\cite{LWang:SQM2026}.}
\label{fig:exotic}
\end{figure}

\subsection{Unstable $^{4}$Li and $^{5}$Li in dense medium at $\snn=3$~GeV}

Low-energy heavy-ion collisions offer an ideal environment for producing light nuclei. The unstable nuclei $^{4}$Li ($\tau\sim30$~fm/$c$, decaying to $^{3}$He$+p$) and $^{5}$Li ($\tau\sim120$~fm/$c$, decaying to $^{4}$He$+p$) provide a qualitatively new way to test the light-nuclei production mechanism. Their lifetimes are comparable to the fireball evolution time, so the coalescence of nucleons into these transient resonances is coupled to final-state scattering. STAR developed a novel method to correct for these final-state interactions, accounting for S-wave and P-wave scattering as well as the resonance channel.

The yields of $^{4}$Li and $^{5}$Li were measured in \AuAu\ collisions at $\snn=3$~GeV and compared with stable light nuclei as a function of mass number (Fig.~\ref{fig:exotic}(a))~\cite{Wu:SQM2026}. The stable species follow the usual exponential penalty-factor scaling expected from coalescence in all four centrality intervals. The unstable species deviate clearly from this baseline. The yield of $^{4}$Li is significantly lower than that of $^{4}$He, and the deviation is larger for the shorter-lived $^{4}$Li than for the longer-lived $^{5}$Li. This lifetime ordering directly reflects the interplay between resonance formation and rescattering in the dense medium.

\subsection{Search for the $K$-$\mu$ muonic atom}

Heavy-ion collisions also copiously produce muonic atoms, Coulomb bound states of a muon and a hadron~\cite{Kapusta:1998fh}. The formation yield is enhanced by electromagnetic coalescence and is sensitive to the yield of soft prompt muons ($\pt\sim0.2$~GeV/$c$) from QGP radiation~\cite{Wang:2024njn}. Such soft muons cannot be measured directly, because the hadron-decay background overwhelms them at low $\pt$. Muonic atoms provide a natural filter: muons from hadron decays are produced too late to bind, so the atom yield selects the prompt component. The atoms are electrically neutral and dissociate in detector material during flight. The kaon and muon tracks can then be reconstructed, but with a characteristic shift in invariant mass, quantified by the residual mass $\delta m \equiv M_{\rm inv}-m_{\mu}-m_{\rm hadron}$, because the dissociation does not occur at the primary vertex.

STAR measured $K$-$\mu$ momentum correlations in the isobar ($^{96}$Ru+$^{96}$Ru and $^{96}$Zr+$^{96}$Zr) collisions at $\snn=200$~GeV, suppressing the residual Coulomb correlation by combining unlike-sign and like-sign pairs. A statistically significant excess consistent with $K\mu$ atom formation is observed in both charge combinations as a function of $\delta m$ (Fig.~\ref{fig:exotic}(b)), with significances of $7.6\sigma$ for $K^{+}\mu^{-}$ and $8.6\sigma$ for $K^{-}\mu^{+}$~\cite{XWang:SQM2026}. Work is ongoing to finalize the acceptance corrections.

\subsection{Coherent $\jpsi$ and $K^{+}K^{-}$ photoproduction in UPCs}

In photonuclear UPC interactions, particle production proceeds through a resonance channel, in which a short-lived vector meson decays to two charged particles ($\gamma+A\rightarrow V+A^{*}\rightarrow h^{+}h^{-}+A^{*}$), or through the non-resonant Drell--S\"oding continuum. In the resonance channel, STAR measured coherent $\jpsi$ photoproduction in the isobar systems through the $e^{+}e^{-}$ decay channel~\cite{Li:SQM2026}. The coherent cross section is directly sensitive to the nuclear gluon distribution. The measured $d\sigma/dy$ falls below the theoretical expectations from STARlight and the impulse approximation, indicating suppression from nuclear shadowing or gluon saturation.

In the non-resonant channel, STAR performed the most precise measurement to date of coherent non-resonant $K^{+}K^{-}$ photoproduction (Fig.~\ref{fig:exotic}(c))~\cite{LWang:SQM2026}. The precision of the pair-$\pt$ spectrum reveals an interference dip at very low $\pt$, arising from the projectile--target ambiguity, and a structure consistent with a diffraction peak encoding the nuclear form factor.

\section{Onset of QGP signatures in \OO\ collisions}
\label{sec:oo}

A central question of the small-system program is at what system size the QGP signatures characteristic of large systems first appear. \OO\ collisions at $\snn=200$~GeV provide an important small-system anchor point. STAR has pursued it with multiple independent probes: jet quenching, quarkonium suppression, thermal radiation, strangeness production, and anisotropic flow.

\begin{figure}[h]
\centering
\includegraphics[width=1.01\linewidth]{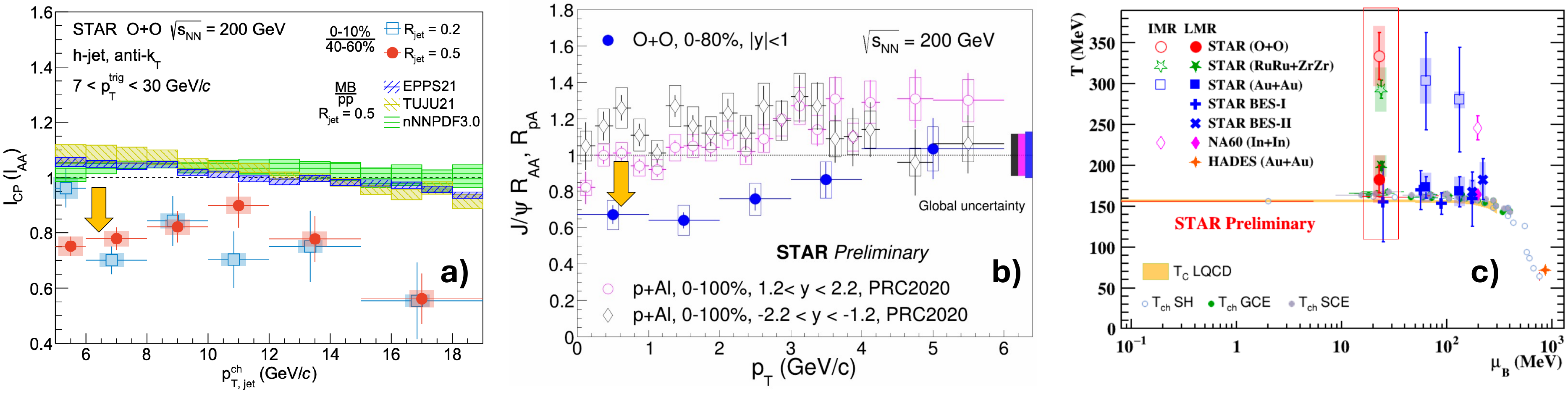}
\caption{Three independent probes of QGP formation in \OO\ collisions at $\snn=200$~GeV. (a) $I_{\rm CP}$ for hadron-jet correlations as a function of recoil-jet $\pt$, showing a $\sim$20\% suppression in central events relative to the nuclear parton distribution function (nPDF) expectations~\cite{STAR:2026nfy}. (b) $\jpsi$ $R_{\rm AA}$ as a function of $\pt$ in \OO\ (filled circles) compared with p+Al $R_{\rm pA}$ measurements (open symbols), which bracket the cold-nuclear-matter baseline~\cite{AZhang:SQM2026}. (c) Effective temperatures extracted from dilepton spectra in the low-mass (LMR, filled) and intermediate-mass (IMR, open) regions as a function of $\mu_{B}$, comparing \OO, isobar, and \AuAu\ systems~\cite{Liu:SQM2026}.}
\label{fig:ooprobes}
\end{figure}

\subsection{Jet quenching, $\jpsi$ suppression, and thermal radiation}

Evidence of jet quenching in \OO\ is observed in away-side hadron-jet and hadron-hadron correlations~\cite{STAR:2026nfy}. The observable $I_{\rm CP}$ is the ratio of per-trigger yields in central relative to peripheral event-activity classes, with event activity defined in forward detectors to reduce selection bias. It shows an approximately 20\% suppression in 0--10\% central collisions (Fig.~\ref{fig:ooprobes}(a)). The magnitude is consistent with partonic energy loss in a small, short-lived medium.

Quarkonium provides a second, independent probe. $\jpsi$ production is sensitive to both hot-medium dissociation/regeneration and cold-nuclear-matter (CNM) effects. In \OO\ collisions, the CNM contribution is constrained using the available $p$+Al measurements, which lie at or slightly above unity. The measured $\jpsi$ $R_{\rm AA}$ shows a clear suppression below unity at low $\pt$ (Fig.~\ref{fig:ooprobes}(b))~\cite{AZhang:SQM2026}. The deviation from the $p$+Al reference is consistent with an additional hot-medium contribution.

Thermal dilepton radiation provides a thermometer of the medium. A smaller system may reach comparable density but with a shorter lifetime, so it is not obvious what effective temperature its radiation reflects. STAR extracted temperatures from the dilepton spectra in both the low-mass region, which is emitted mostly near the phase transition, and the intermediate-mass region, which is dominated by early QGP radiation (Fig.~\ref{fig:ooprobes}(c))~\cite{Liu:SQM2026}. Both temperatures are consistent with those of the larger isobar system. The temperature reached in \OO\ is therefore as high as in large systems, consistent with thermal radiation from a QGP medium.

Strangeness production completes the picture. Strangeness enhancement versus system size is a classic QGP signature, and at RHIC the onset region between $pp$ and large systems was previously unpopulated. New STAR measurements of strange-hadron yields ($\Lambda$, $\Xi$, $\Omega$, $\phi$) versus $\npart$ in \OO\ fill this gap and reveal a smooth transition from small to large systems~\cite{Ponce:SQM2026}.

\subsection{Anisotropic flow: disentangling nucleonic and subnucleonic fluctuations}

For flow observables, the goal is to establish that the signal is driven by initial-state geometry through the QGP response. The earlier small-system scan in asymmetric systems ($p/d/^{3}$He+Au) left a substantial ambiguity on this point~\cite{STAR:2022pfn,STAR:2023wmd}. The predicted eccentricity ordering depends on both the nucleon configurations and the subnucleonic fluctuations. In a nucleon Glauber model, the ordering of $\varepsilon_{2}$ across the three systems reflects the intrinsic two- and three-nucleon geometry of the projectile. Once quark substructure is included, both the magnitudes and the ordering of $\varepsilon_{2}$ and $\varepsilon_{3}$ change dramatically~\cite{Welsh:2016siu,Huang:2025cjm}. The asymmetric-system data alone could not separate the two sources.

STAR's strategy is to constrain the subnucleonic component by comparing \dAu\ with \OO~\cite{STAR:2025ivi}. The nucleon configurations of the deuteron and of $^{16}$O are both well constrained by ab initio nuclear-structure calculations, so any residual model freedom resides in the subnucleonic sector. At the same multiplicity, the measured $v_{2}$ in \dAu\ is much larger than in \OO, while $v_{3}$ is comparable. This ordering follows from the eccentricities: the elongated deuteron drives a large intrinsic $\varepsilon_{2}$ in \dAu, whereas the triangularity is fluctuation-driven in both systems. Scaling the flow coefficients by the eccentricity, $k_{n}=v_{n}/\varepsilon_{n}$, isolates the hydrodynamic response, which is expected to depend primarily on multiplicity within the hydrodynamic framework. The \dAu\ and \OO\ data indeed collapse onto a common curve, and the collapse clearly improves when the eccentricities include subnucleonic fluctuations. The combined \dAu\ and \OO\ data are thus directly sensitive to substructure in the initial state.

\begin{figure}[t]
\centering
\includegraphics[width=0.8\linewidth]{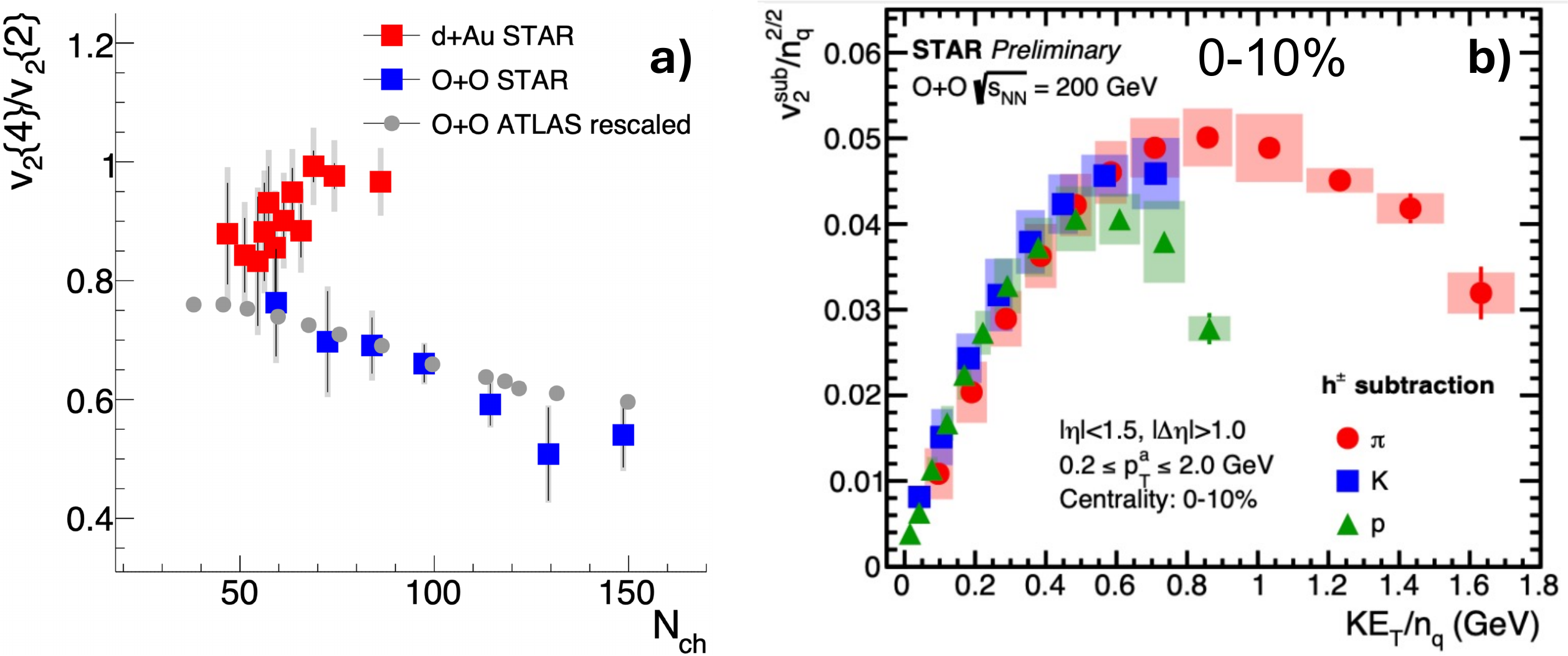}
\caption{(a) Ratio $v_{2}\{4\}/v_{2}\{2\}$ as a function of $\nch$ in \dAu\ (red squares) and \OO\ (blue squares) collisions at $\snn=200$~GeV~\cite{STAR:2025ivi}, compared with ATLAS \OO\ data at $\snn=5.36$~TeV (grey circles), re-matched to the STAR centrality definition~\cite{ATLAS:2025nnt}. (b) NCQ-scaled elliptic flow $v_{2}/n_{q}$ versus $KE_{\rm T}/n_{q}$ for $\pi$, $K$, and $p$ in 0--10\% central \OO\ collisions~\cite{Paul:SQM2026}. All the data are obtained after nonflow subtraction.}
\label{fig:ooflow}
\end{figure}

The elliptic-flow fluctuations sharpen this picture. The ratio $v_{2}\{4\}/v_{2}\{2\}$ approaches unity when the ellipticity has a dominant intrinsic component, and it decreases as fluctuations grow. In \dAu\ the prolate deuteron guarantees a large intrinsic $\varepsilon_{2}$ in every central event. In \OO\ the ellipticity is almost entirely fluctuation-driven. The data bear this out: $v_{2}\{4\}/v_{2}\{2\}\approx\varepsilon_{2}\{4\}/\varepsilon_{2}\{2\}\approx0.9$ in \dAu\ versus $\sim$0.6 in \OO\ (Fig.~\ref{fig:ooflow}(a)). The ATLAS \OO\ measurement at the LHC~\cite{ATLAS:2025nnt}, re-matched to the STAR centrality definition, agrees with the RHIC data. Within current uncertainties, the initial-geometry fluctuations show little beam-energy dependence, despite a factor of ten increase in $\snn$ and a factor of three in multiplicity.

Identified-particle flow provides another consistency check on collectivity. The elliptic flow of $\pi$, $K$, and $p$ in central \OO\ collisions exhibits the characteristic mass-dependent hierarchy versus $\pt$. After number-of-constituent-quark (NCQ) scaling of both axes, these species collapse onto a common curve at low $\pt$ (Fig.~\ref{fig:ooflow}(b)), supporting the collective, partonic origin of the signal~\cite{Paul:SQM2026}.

\begin{figure}[t]
\centering
\includegraphics[width=0.8\linewidth]{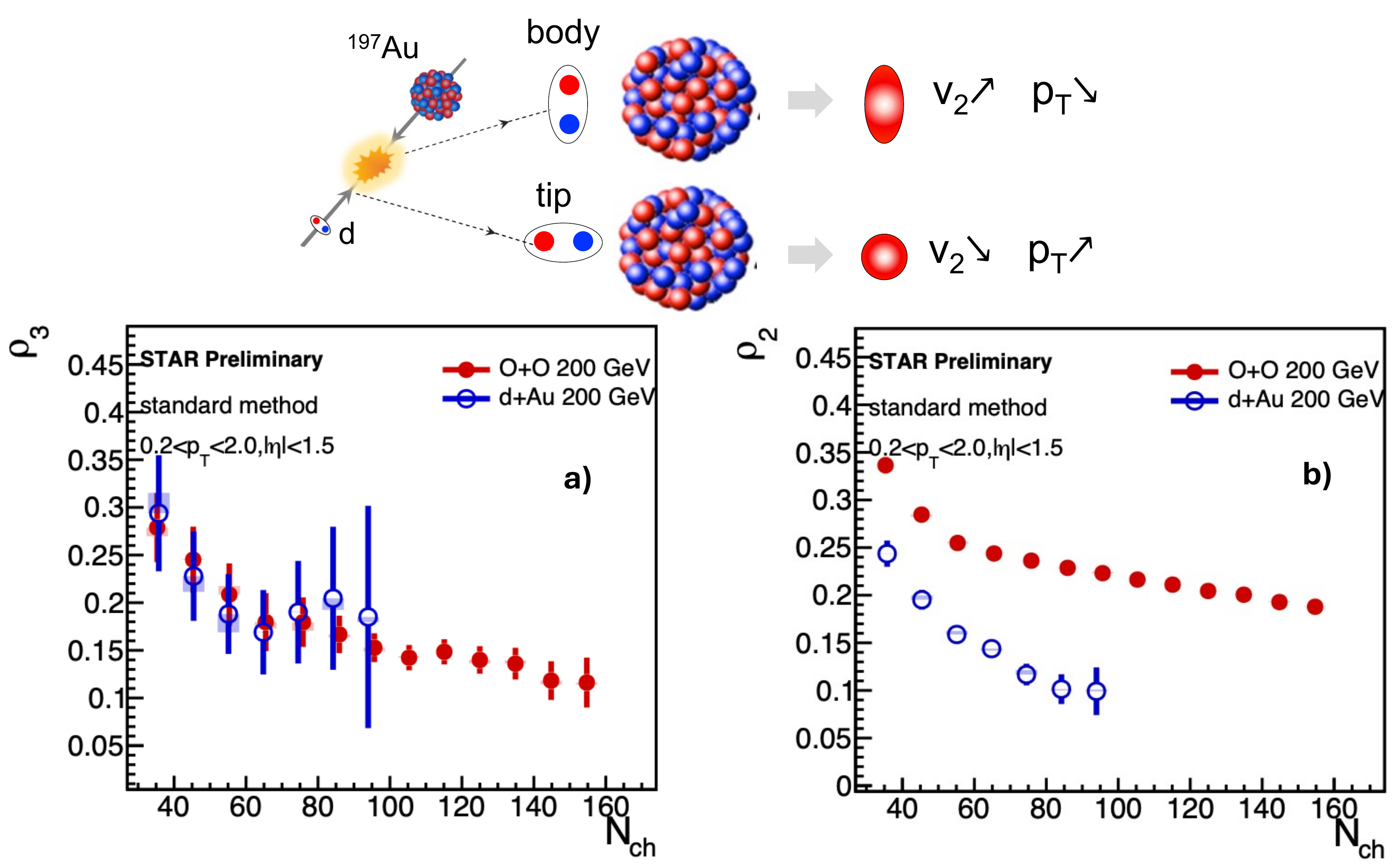}
\caption{(Top) The cartoon illustration of the anti-correlation between $v_{2}^{2}$ and $\delta\pt=[\pt]-\langle\pt\rangle$ in \dAu\ collisions. (Bottom) Pearson correlation coefficients (a) $\rho_{3}$ and (b) $\rho_{2}$ between $v_{n}^{2}$ and $\delta\pt$ as a function of $\nch$ in \dAu\ and \OO\ collisions at $\snn=200$~GeV. The strong suppression of $\rho_{2}$ in \dAu, absent in $\rho_{3}$, reflects the QGP response to the prolate deuteron shape.}
\label{fig:deuteron}
\end{figure}

The prolate shape of the deuteron leaves one more distinctive imprint in multi-particle correlations. The orientation of the deuteron at impact, body-on versus tip-on, creates an anti-correlation between $v_{2}$ and the mean transverse momentum $\mpt$. Body-on collisions produce a large elliptic overlap area: large $v_{2}$ and small $\mpt$. Tip-on collisions give the reverse (top of Fig.~\ref{fig:deuteron}). The same mechanism is established in U+U collisions, where it reduces the $v_{2}$--$\pt$ covariance according to $\langle v_{2}^{2}\delta\pt\rangle = a - b\beta_{2}^{3}$~\cite{Jia:2021qyu,STAR:2024eky}. STAR quantified it via the normalized Pearson coefficients $\rho_{n} = \langle v_{n}^{2}\delta\pt\rangle/(\langle v_{n}^{2}\rangle\sqrt{\langle(\delta\pt)^2\rangle})$ for $n=2$ and 3 (Fig.~\ref{fig:deuteron}(a)--(b)). At the same multiplicity, $\rho_{2}$ in \dAu\ is strongly suppressed relative to \OO. In contrast, $\rho_{3}$ is consistent between the two systems, as expected since the triangularity carries no information about the deuteron shape. This is precisely the expected pattern for a QGP responding hydrodynamically to the prolate deuteron geometry.

\section{Radial flow and its fluctuations}
\label{sec:radial}

Radial flow reflects the hydrodynamic response to the initial overlap area. For a given total energy, a large source has a small pressure gradient and yields a small $\mpt$; a compact source yields a large $\mpt$. The overlap size $R$ fluctuates event-by-event within a centrality class, so the mean $\pt$ fluctuates as well. Hydrodynamic simulations show a tight linear anti-correlation between the two, $\delta\pt/\langle[\pt]\rangle \approx -\delta R/\langle R\rangle$. Three observables characterize this response: the mean transverse momentum $\langle[\pt]\rangle$; its integral radial-flow fluctuation
\begin{equation}
v_{0,{\rm int}} \equiv \frac{\sqrt{\langle(\delta[\pt])^{2}\rangle}}{\langle[\pt]\rangle};
\end{equation}
and the $\pt$-differential fluctuation $v_{0}(\pt)$, obtained from the covariance of the local spectrum with the global event-wise $[\pt]$,
\begin{equation}
v_{0}(\pt)\,v_{0,{\rm int}} = \frac{\langle \delta n(\pt)\,\delta[\pt]\rangle}{\langle n(\pt)\rangle\,\langle[\pt]\rangle}.
\end{equation}
The differential $v_{0}(\pt)$ captures the coherent pivoting of the spectrum. In an event with above-average $[\pt]$, the entire spectrum hardens: the yield is depleted at low $\pt$ and enhanced at high $\pt$. An event with below-average $[\pt]$ shows the opposite pattern.

\subsection{Constraining speed-of-sound in ultra-central collisions}

In ultra-central collisions, $\mpt$ rises with $\nch$. This rise has been related to temperature fluctuations at fixed volume through a thermodynamic relation, giving direct access to the speed of sound, $\cs \approx d(\ln\mpt)/d(\ln\nch)$~\cite{Gardim:2019xjs}. Recent model studies showed, however, that this relation is biased by the choice of centrality estimator: the extracted slope depends dramatically on the $\pt$ threshold used to count $\nch$~\cite{Nijs:2023bzv,ALICE:2025rtg}. An alternative extraction relates the joint $(\nch,\mpt)$ fluctuations to $\cs$ with an explicit event-by-event bias term, which can be eliminated by imposing Gaussianity of the initial-state fluctuations~\cite{Mu:2025gtr}. STAR combines the two methods within the Trajectum framework. The fluctuation method fixes $\cs$; the $\pt$ threshold for $\nch$ is then varied until the slope method agrees. This procedure selects $p_{\rm T,min}=0.225$~GeV/$c$ as the threshold that removes the bias in the STAR acceptance~\cite{Broodo:SQM2026}.

\begin{figure}[t]
\centering
\includegraphics[width=0.8\linewidth]{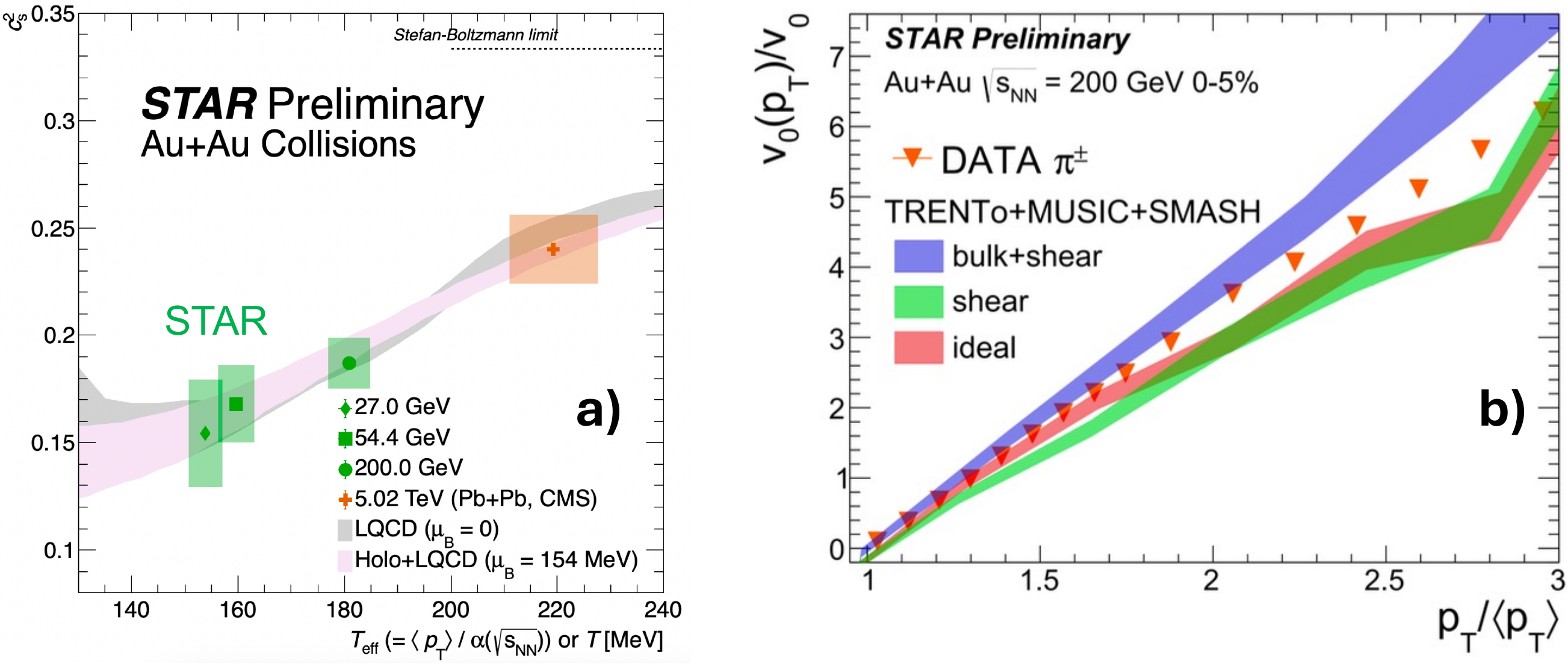}
\caption{(a) Estimate of speed of sound squared as a function of the effective temperature $T_{\rm eff}$ extracted from ultra-central \AuAu\ collisions at $\snn=27$, 54.4, and 200~GeV~\cite{Broodo:SQM2026}, compared with the CMS Pb+Pb result at 5.02~TeV, lattice QCD at $\mu_{B}=0$, and a holographic QCD calculation at $\mu_{B}=154$~MeV. (b) Ratio $v_{0}(\pt)/v_{0}$ in 0--5\% \AuAu\ collisions at $\snn=200$~GeV for pions, normalized to $\pt/\mpt$, compared with TRENTo+MUSIC+SMASH calculations for ideal, shear-only, and bulk+shear viscous hydrodynamics~\cite{ZWang:SQM2026}.}
\label{fig:cs2}
\end{figure}

With the bias controlled, STAR obtained a preliminary estimate of $\cs$ at three energies: $\cs = 0.154\pm0.026$ at $\snn=27$~GeV, $0.168\pm0.018$ at 54.4~GeV, and $0.187\pm0.013$ at 200~GeV. Each value is paired with an effective temperature obtained from the fully extrapolated $\mpt$ via $T_{\rm eff}=\alpha\mpt/3$, where $\alpha$ is close to unity in the Trajectum hydrodynamic model. The result (Fig.~\ref{fig:cs2}(a)) provides an estimate of the temperature dependence of the speed of sound in the crossover region~\cite{Broodo:SQM2026}. The estimates are compatible with lattice-QCD calculations within the current uncertainties. Together with the CMS Pb+Pb point at 5.02~TeV, it maps the rise of $\cs$ over $T_{\rm eff}\approx150$--220~MeV.

\subsection{Integral and differential $v_{0}$}

The integral $v_{0}$ was measured for charged hadrons and for identified $\pi$, $K$, $p$ from \OO\ to \AuAu\ at $\snn=200$~GeV~\cite{ZWang:SQM2026}. The fluctuation signal is larger for heavier species, as expected since radial flow imparts a larger momentum boost to heavier particles. At matched multiplicity, the \OO\ data fall on top of the \AuAu\ data for each species. Across systems, $v_{0}$ follows an approximate power-law scaling, $v_{0}\sim1/\sqrt{\nch}$, as expected if the fireball is built from independent sources.

The differential $v_{0}(\pt)$, normalized by the integral $v_{0}$ to remove the overall radial-flow fluctuation, collapses onto a universal shape at low $\pt$. The collapse holds across all centralities and across systems from \OO\ to \AuAu; deviations appear only at higher $\pt$. The intermediate-$\pt$ region around $\pt\sim2$~GeV is particularly sensitive to viscous effects. Comparison of the pion data with TRENTo+MUSIC+SMASH calculations shows that ideal and shear-only evolutions are nearly indistinguishable there. Including bulk viscosity displaces the curve sizably (Fig.~\ref{fig:cs2}(b)). The ratio $v_{0}(\pt)/v_0$ therefore provides a promising constraint on the QGP bulk viscosity.

The same observable has been extended to strange hadrons ($K_{S}^{0}$, $\phi$, $\Lambda$, $\Xi$) across centralities at 200~GeV and across beam energies down to 11.5~GeV~\cite{Kong:SQM2026}. NCQ scaling of $v_{0}(\pt)$ holds within uncertainties across this broad centrality and energy range, and the signal changes only weakly with $\snn$. This mirrors the NCQ scaling of anisotropic flow and points to a common partonic origin of the radial expansion. This rich dataset awaits detailed model confrontation.

\section{Polarization and spin correlations}
\label{sec:spin}

\begin{figure}[t]
\centering
\includegraphics[width=1\linewidth]{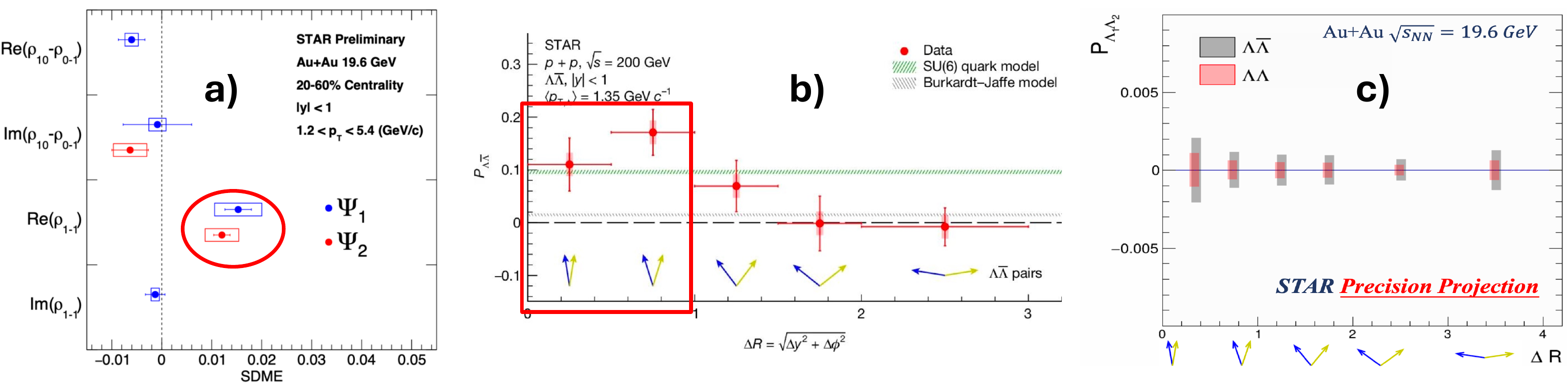}
\caption{(a) Spin-density matrix elements of the $\phi$ meson measured via the two-dimensional $(\theta^{*},\beta)$ distribution of daughter kaons in 20--60\% \AuAu\ collisions at $\snn=19.6$~GeV, using first-order ($\Psi_{1}$) and second-order ($\Psi_{2}$) event planes. The diagonal $\rho_{00}-1/3$ and the off-diagonal ${\rm Re}(\rho_{1-1})$ are both significantly nonzero~\cite{Wilks:SQM2026}. (b) $\lam\lambar$ spin correlation $P_{\lam\lambar}$ versus pair separation $\Delta R$ in $pp$ collisions at $\sqrt{s}=200$~GeV~\cite{STAR:2025njp}, compared with the SU(6) quark model and the Burkardt--Jaffe model. (c) STAR precision projection for the $\lam\lam$ and $\lam\lambar$ spin-correlation measurement in \AuAu\ collisions at $\snn=19.6$~GeV.}
\label{fig:sdme}
\end{figure}

The $\phi$-meson spin alignment is a unique probe of QGP vorticity and, more broadly, of spin dynamics in the medium. The spin state of this vector meson is described by a $3\times3$ spin-density matrix,
\begin{equation}
\rho=\left(\begin{array}{ccc}
\rho_{-1-1} & \rho_{-1,0} & \rho_{-11} \\
\rho_{0-1} & \rho_{00} & \rho_{01} \\
\rho_{1-1} & \rho_{10} & \rho_{11}
\end{array}\right),
\end{equation}
whose elements are fully encoded in the angular distribution of the daughter kaon in the $\phi$ rest frame. Previous measurements assumed vanishing off-diagonal elements and accessed only $\rho_{00}$ through the one-dimensional polar-angle ($\theta^{*}$) distribution, which integrates out the azimuthal information. STAR has now extended the measurement to the full two-dimensional $(\theta^{*},\beta)$ distribution, giving access to the off-diagonal elements for the first time.

The measurement, performed in 20--60\% \AuAu\ collisions at $\snn=19.6$~GeV with both first- and second-order event planes ($\Psi_1$ and $\Psi_2$), is the first in \AuAu\ with the full two-dimensional correction (Fig.~\ref{fig:sdme}(a))~\cite{Wilks:SQM2026}. The results from the two event planes agree, providing an internal consistency check. The key finding is the first evidence of a nonzero off-diagonal element: ${\rm Re}(\rho_{1-1})$ is positive at the level of one to two percent, comparable in magnitude to the diagonal $\rho_{00}-1/3$. The combined statistical and systematic significance of ${\rm Re}(\rho_{1-1})$ is 2.8 and 3.2 standard deviations for $\Psi_1$ and $\Psi_2$, respectively. Since the $\phi$ is an $s\bar{s}$ bound state, both global spin alignment and local $s$--$\bar{s}$ spin correlations contribute to the density matrix. The large off-diagonal signal points to a nonzero spin--spin correlation component.

Direct evidence for such $s\bar{s}$ spin correlations already exists in $pp$ collisions: STAR recently published the measurement of $\lam\lambar$ spin correlations at $\sqrt{s}=200$~GeV~\cite{STAR:2025njp}, which probes the spin correlation of the parent $s\bar{s}$ pair. The $\lam$ decay is self-analyzing: the proton in $\lam\rightarrow p\pi^{-}$ is emitted preferentially along the spin direction, so the pair spin correlation is read off from the joint decay angular distribution. At small angular separation the correlation signal is approximately eighteen percent (Fig.~\ref{fig:sdme}(b)). This is a strikingly large value. It is consistent with the $J^{PC}=0^{++}$ spin correlations of the QCD vacuum being transferred nearly intact to the final state, at the level expected in the SU(6) quark model. The signal vanishes when the pairs are widely separated in angle, a pattern consistent with decoherence of the pair.

The natural next question is what happens to the QCD vacuum at finite temperature, that is, in the QGP. At high temperature the vacuum condensates are expected to melt, so the correlation could be reduced or erased in the medium. The corresponding $\lam\lambar$ measurement in \AuAu\ collisions at BES energies is ongoing, with the projection shown for $\snn=19.6$~GeV, where the expected precision is better than 0.1\% (Fig.~\ref{fig:sdme}(c))~\cite{Fu:SQM2026}. If the signal survives in the medium, STAR will measure it decisively. This is a result to look forward to at the next SQM.

\section{Summary}
\label{sec:summary}

STAR's rich datasets, with more than forty species combinations and energies accumulated over twenty-five years of RHIC operation, are a gold mine for a diverse physics program. This contribution highlighted four threads of recent progress. First, the first observations of the unstable nuclei $^{4}$Li and $^{5}$Li, a statistically significant excess consistent with $K\mu$ atom formation, and precision coherent photoproduction measurements of $\jpsi$ and non-resonant $K^{+}K^{-}$ in UPCs. Second, strong evidence for QGP formation in central \OO\ collisions from multiple independent probes: jet quenching, $\jpsi$ suppression beyond the cold-nuclear-matter baseline, thermal dilepton radiation, strangeness enhancement, and anisotropic flow. The \dAu/\OO\ comparison further constrains subnucleonic fluctuations in the initial state. Third, new radial-flow observables that deliver a preliminary estimate of the temperature dependence of the speed of sound, compatible with lattice QCD, and provide a promising constraint on the bulk viscosity. Fourth, the first measurements of off-diagonal $\phi$-meson spin-density-matrix elements and of $\lam\lambar$ spin correlations, opening a new window on the spin structure of the QCD vacuum at zero and finite temperature. More discoveries are to come in the decades ahead.

\section*{Acknowledgment}
This work is supported in part by the U.S. Department of Energy, Office of Science, Office of Nuclear Physics, Grant Number DE-SC0024602.

\bibliography{ref}{}

\begin{thebibliography}{10}
\expandafter\ifx\csname url\endcsname\relax
  \def\url#1{\texttt{#1}}\fi
\expandafter\ifx\csname urlprefix\endcsname\relax\def\urlprefix{URL }\fi
\expandafter\ifx\csname href\endcsname\relax
  \def\href#1#2{#2} \def\path#1{#1}\fi

\bibitem{Wu:SQM2026}
{J. Wu (for the STAR Collaboration)}, these proceedings.

\bibitem{XWang:SQM2026}
{X. Wang (for the STAR Collaboration)}, these proceedings.

\bibitem{LWang:SQM2026}
{L. Wang (for the STAR Collaboration)}, these proceedings.

\bibitem{Kapusta:1998fh}
J.~I. Kapusta, A.~Mocsy, Phys. Rev. C 59 (1999) 2937--2940.
\newblock \href {http://arxiv.org/abs/nucl-th/9812013}
  {\path{arXiv:nucl-th/9812013}}, \href
  {https://doi.org/10.1103/PhysRevC.59.2937}
  {\path{doi:10.1103/PhysRevC.59.2937}}.

\bibitem{Wang:2024njn}
X.~Wang, F.~Geurts, Z.~Tang, K.~Xin, Z.~Xu, Y.~Zhang, L.~Zhou, Phys. Lett. B
  861 (2025) 139242.
\newblock \href {http://arxiv.org/abs/2410.06518} {\path{arXiv:2410.06518}},
  \href {https://doi.org/10.1016/j.physletb.2025.139242}
  {\path{doi:10.1016/j.physletb.2025.139242}}.

\bibitem{Li:SQM2026}
{Z. Li (for the STAR Collaboration)}, these proceedings.

\bibitem{STAR:2026nfy}
{STAR Collaboration}, arXiv:2604.13935 (2026).

\bibitem{AZhang:SQM2026}
{A. Zhang (for the STAR Collaboration)}, these proceedings.

\bibitem{Liu:SQM2026}
{Z. Liu (for the STAR Collaboration)}, these proceedings.

\bibitem{Ponce:SQM2026}
{I. Ponce (for the STAR Collaboration)}, these proceedings.

\bibitem{STAR:2022pfn}
M.~I. Abdulhamid, et~al., Phys. Rev. Lett. 130~(24) (2023) 242301.
\newblock \href {http://arxiv.org/abs/2210.11352} {\path{arXiv:2210.11352}},
  \href {https://doi.org/10.1103/PhysRevLett.130.242301}
  {\path{doi:10.1103/PhysRevLett.130.242301}}.

\bibitem{STAR:2023wmd}
M.~I. Abdulhamid, et~al., Phys. Rev. C 110~(6) (2024) 064902.
\newblock \href {http://arxiv.org/abs/2312.07464} {\path{arXiv:2312.07464}},
  \href {https://doi.org/10.1103/PhysRevC.110.064902}
  {\path{doi:10.1103/PhysRevC.110.064902}}.

\bibitem{Welsh:2016siu}
K.~Welsh, J.~Singer, U.~W. Heinz, Phys. Rev. C 94~(2) (2016) 024919.
\newblock \href {http://arxiv.org/abs/1605.09418} {\path{arXiv:1605.09418}},
  \href {https://doi.org/10.1103/PhysRevC.94.024919}
  {\path{doi:10.1103/PhysRevC.94.024919}}.

\bibitem{Huang:2025cjm}
S.~Huang, J.~Jia, C.~Zhang, Phys. Lett. B 870 (2025) 139926.
\newblock \href {http://arxiv.org/abs/2507.16162} {\path{arXiv:2507.16162}},
  \href {https://doi.org/10.1016/j.physletb.2025.139926}
  {\path{doi:10.1016/j.physletb.2025.139926}}.

\bibitem{STAR:2025ivi}
{STAR Collaboration}, arXiv:2510.19645 (2025).

\bibitem{ATLAS:2025nnt}
G.~Aad, et~al., Phys. Rev. C 113~(4) (2026) 045205.
\newblock \href {http://arxiv.org/abs/2509.05171} {\path{arXiv:2509.05171}},
  \href {https://doi.org/10.1103/xqxz-8bhf} {\path{doi:10.1103/xqxz-8bhf}}.

\bibitem{Paul:SQM2026}
{S. Paul (for the STAR Collaboration)}, these proceedings.

\bibitem{Jia:2021qyu}
J.~Jia, Phys. Rev. C 105~(1) (2022) 014905.
\newblock \href {https://doi.org/10.1103/PhysRevC.105.014905}
  {\path{doi:10.1103/PhysRevC.105.014905}}.

\bibitem{STAR:2024eky}
M.~I. Abdulhamid, et~al., Nature 635~(8037) (2024) 67--72.
\newblock \href {https://doi.org/10.1038/s41586-024-08097-2}
  {\path{doi:10.1038/s41586-024-08097-2}}.

\bibitem{Gardim:2019xjs}
F.~G. Gardim, G.~Giacalone, M.~Luzum, J.-Y. Ollitrault, Nature Phys. 16~(6)
  (2020) 615--619.
\newblock \href {http://arxiv.org/abs/1908.09728} {\path{arXiv:1908.09728}},
  \href {https://doi.org/10.1038/s41567-020-0846-4}
  {\path{doi:10.1038/s41567-020-0846-4}}.

\bibitem{Nijs:2023bzv}
G.~Nijs, W.~van~der Schee, Phys. Lett. B 853 (2024) 138636.
\newblock \href {http://arxiv.org/abs/2312.04623} {\path{arXiv:2312.04623}},
  \href {https://doi.org/10.1016/j.physletb.2024.138636}
  {\path{doi:10.1016/j.physletb.2024.138636}}.

\bibitem{ALICE:2025rtg}
I.~J. Abualrob, et~al., JHEP 11 (2025) 076.
\newblock \href {http://arxiv.org/abs/2506.10394} {\path{arXiv:2506.10394}},
  \href {https://doi.org/10.1007/JHEP11(2025)076}
  {\path{doi:10.1007/JHEP11(2025)076}}.

\bibitem{Mu:2025gtr}
Y.-S. Mu, J.-A. Sun, L.~Yan, X.-G. Huang, Phys. Rev. Lett. 135~(16) (2025)
  162301.
\newblock \href {http://arxiv.org/abs/2501.02777} {\path{arXiv:2501.02777}},
  \href {https://doi.org/10.1103/skhj-cj9p} {\path{doi:10.1103/skhj-cj9p}}.

\bibitem{Broodo:SQM2026}
{C. Broodo (for the STAR Collaboration)}, these proceedings.

\bibitem{ZWang:SQM2026}
{Z. Wang (for the STAR Collaboration)}, these proceedings.

\bibitem{Kong:SQM2026}
{Y. Kong (for the STAR Collaboration)}, these proceedings.

\bibitem{Wilks:SQM2026}
{G. Wilks (for the STAR Collaboration)}, these proceedings.

\bibitem{STAR:2025njp}
B.~E. Aboona, et~al., Nature 650~(8100) (2026) 65--71.
\newblock \href {http://arxiv.org/abs/2506.05499} {\path{arXiv:2506.05499}},
  \href {https://doi.org/10.1038/s41586-025-09920-0}
  {\path{doi:10.1038/s41586-025-09920-0}}.

\bibitem{Fu:SQM2026}
{T. Fu (for the STAR Collaboration)}, these proceedings.

\end{thebibliography}
\bibliographystyle{elsarticle-num}
\end{document}